# An Effective Activation Method for Industrially Produced TiFeMn Powder for Hydrogen Storage

David Michael Dreistadt [a], Thi-Thu Le [a], Giovanni Capurso [a,1*], José M. Bellosta von Colbe [a], Archa Santhosh [a], Claudio Pistidda [a], Nico Scharnagl [b], Henry Ovri [c], Chiara Milanese [d], Paul Jerabek [a], Thomas Klassen [a,e], Julian Jepsen [a,e]

[a] Institute of Hydrogen Technology, Helmholtz-Zentrum hereon GmbH, Max-Planck-Straße 1, 21502 Geesthacht, Germany
[b] Institute of Surface Science, Helmholtz-Zentrum hereon GmbH, Max-Planck-Straße 1, 21502 Geesthacht, Germany
[c] Institute of Materials Mechanics, Helmholtz-Zentrum hereon GmbH, Max-Planck-Straße 1, 21502 Geesthacht, Germany
[d] Pavia Hydrogen Lab, C.S.G.I. & Chemistry Department, University of Pavia, viale Taramelli 16, 27100 Pavia, Italy
[e] Helmut Schmidt University, Holstenhofweg 85, 22043 Hamburg, Germany

* Corresponding author: giovanni.capurso@hereon.de

## Abstract

This work proposes an effective thermal activation method with low technical effort for industrially produced titanium-iron-manganese powders (TiFeMn) for hydrogen storage. In this context, the influence of temperature and particle size of TiFeMn on the activation process is systematically studied. The results obtained from this investigation suggest that the activation of the TiFeMn material at temperatures as low as 50 °C is already possible, with a combination of "Dynamic" and "Static" routines, and that an increase to 90 °C strongly reduces the incubation time for activation, *i.e.* the incubation time of the sample with the two routines at 90 °C is about 0.84 h, while ~ 277 h is required for the sample treated at 50 °C in both "Dynamic" and "Static" sequences. Selecting TiFeMn particles of larger size also leads to significant improvements in the activation performance of the investigated material. The proposed activation routine makes it possible to overcome the oxide layer existing on the compound surface, which acts as a diffusion barrier for the hydrogen atoms. This activation method induces further cracks and defects in the powder granules, generating new surfaces for hydrogen absorption with greater frequency, and thus leading to faster sorption kinetics in the subsequent absorption-desorption cycles.

---

[1] Present address: Polytechnic Department of Engineering and Architecture, University of Udine, via del Cotonificio 108, 33100 Udine, Italy

## 1. Introduction

The high energy demand of modern industrialized nations has a significant negative impact on the environment. This is primarily due to the use of fossil fuels as energy sources. The environmental problems caused by this are forcing politicians and scientists to search for alternative and clean energy sources. In this context, hydrogen appears to be an extremely appealing energy vector, due to its advantages, such as its energy density per unit of mass, which is about three times higher than that of gasoline (141.86 MJ/kg *vs.* 46.4 MJ/kg), despite its energy density per unit of volume being lower than that of gasoline (10.044 MJ/L for liquid $H_2$ *vs.* 34.2 MJ/L) [1-3]. In comparison to the physical storage of hydrogen in gaseous and liquid forms, *i.e.* compressed gas in pressure vessels or liquefied in cryogenic tanks, hydrogen storage in solid-state materials is most likely to represent a suitable solution, due to its higher volumetric density, low loading pressure, moderate temperature of operation, and increased safety level due to the bonding of the hydrogen by the storage material [4, 5].

Since the first reports on the possibility of titanium-iron alloys (TiFe) to store and release hydrogen at near room temperature and low pressure for practical applications [6-8], TiFe based intermetallic alloys have been widely investigated as reversible hydrogen storage materials, particularly for industrial applications [9-12]. The main advantage of TiFe is that it can operate at temperatures below 100 °C and pressures below 50 bar $H_2$, which possibly reduce the technical effort and costs due to simpler tank construction methods [13]. Besides, the high volumetric hydrogen storage capacity (0.096 kg $H_2$/L system [1, 2]), the non-toxicity, and the cost-competitiveness make this alloy a potential candidate for energy storage applications (*i.e.* coupling with fuel cells in stationary hydrogen storage applications [14]). However, the activation taking place at a high temperature of up to 397 °C is the key issue that limits the practical use of the TiFe based systems [15-17]. Reilly *et al.* [6] in 1974 first reported the activation procedure of TiFe for hydrogen absorption by repeated exposure to ~ 65 bar of hydrogen at elevated temperatures as ~ 400 °C. However, when in contact with air, the activated materials are quickly contaminated, leading to the termination of the following desorption-adsorption processes. Since then, many efforts to improve the activation in TiFe or to improve the synthesis methods without requiring an activation process for TiFe based materials have been made. The most common way is the partial substitution of the constituent elements of TiFe with a third transition metal, *i.e.* manganese (Mn), cerium (Ce), nickel (Ni), cobalt (Co), aluminum (Al), vanadium (V), yttrium (Y), palladium (Pd), chromium (Cr), and zirconium (Zr) or the addition of further elements [18-30] The presence of substitutional impurities in TiFe promotes the diffusion of hydrogen in the bulk that makes the activation possible to take place at lower pressures and temperatures. On the contrary, this method often leads to a loss of the hydrogen storage capacity, particularly for the substitution of Fe with Al and Ni [22, 23]. The treatment based on the introduction of mechanical forces is another effective way for activating TiFe: ball-milling and cryo-milling [31, 32], cold rolling [33], groove rolling [31, 34], equal channel angular pressing [31], severe plastic deformation (SPD) through high-pressure torsion (HPT) [15, 28, 31, 34-36], and the most recently proposed shake or stir method [37]. These mechanical treatments change the microstructure of TiFe and mainly generate unreacted new TiFe surfaces, which help to accelerate the initial activation and subsequently enhance the hydrogenation properties of the material. Besides, surface modification can also be considered for improving the rate of the initial activation of TiFe, in which the surface of the material to be activated is treated so that the activation can take place under technically more favorable conditions. In addition, the catalytic effect of the particle surface properties can improve the dissociative chemisorption of hydrogen molecules [38]. This is especially important due to the fact that the hydrogenation properties of TiFe based compounds are very sensitive to impurities like oxygen and such methods can reduce the negative effect of this. As an example of the deposition of metals on the surface [38, 39] a metal-organic chemical vapor deposition (CVD) technique was employed to deposit metallic palladium on TiFe surface or argon- and hydrogen-ions [40] were implemented in the material's surface structure to improve the first hydrogenation. Nevertheless, the main effect of the activation procedure is to remove or weaken the surface oxide layers, and to create new unreacted surfaces by multiple hydrogen absorption-desorption cycles, with the purpose of promoting hydrogen diffusion [41].

This work aims at refining an activation method that requires the lowest possible technical effort and thus reduces the final cost for the use of industrially-produced powder, hereafter labeled "TiFeMn". Moreover, aiming at operating the hydrogen storage system at low pressure and temperatures, the method should allow the material activation to take place directly in the storage vessel. To keep this research closer to a practical application, TiFeMn alloy obtained from an industrial manufacturer was used as a hydrogen storage material. In the work here described, a thermal treatment was used to activate the TiFeMn powders. This approach is close to the thermochemical treatments reported in the literature [42, 43], with the main difference that, in this case, the focus is on temperatures below 100 °C. Additionally, the presence of an oxide layer on the surface of the investigated alloy is expected, thus preventing the hydrogen from penetrating the material so that activation at room temperature is more difficult. Hence, the influence of temperature and particle size on the activation process of the TiFeMn material was also investigated and the results were discussed in detail.

## 2. Material and methods

*Material:* TiFeMn alloy in powdery form was received from the company GKN Powder Metallurgy with particle sizes ranging up to 600 μm. The TiFeMn powder was produced by melting its constituent elements (Ti, Fe, and Mn) into ingots and then crushing them mechanically. The subsequent storage was done under an air atmosphere. The composition of the material is evaluated as 40-60 wt. % Ti, 40-60 wt. % Fe and 4-6 wt. % Mn [32].

The *activation routine* was carried out using a custom-made Sieverts' apparatus. This routine consists of two parts: "Dynamic" and "Static", as illustrated in Fig. 1. For each activation, 6 g of the as-received TiFeMn sample were loaded into the sample holder and then connected to the Sieverts' apparatus. The sample holder was evacuated under a dynamic rotary vacuum (0.01 bar) for 250 s and then refilled with hydrogen (40 bar $H_2$) for 60 s. This procedure "Dynamic" was repeated 10 times before introducing 40 bar $H_2$ until the material is activated ("Static"). An attempt of 9 experiments at different "Static" temperatures (50 °C, 70 °C, and 90 °C) and at different "Dynamic" temperatures (50 °C, 70 °C and 90 °C) was made. More specifically, following each of the "Static" temperatures, the three different "Dynamic" temperatures were applied. The temperatures of "Dynamic" and "Static" were monitored to study how the temperature affects the activation of the material.

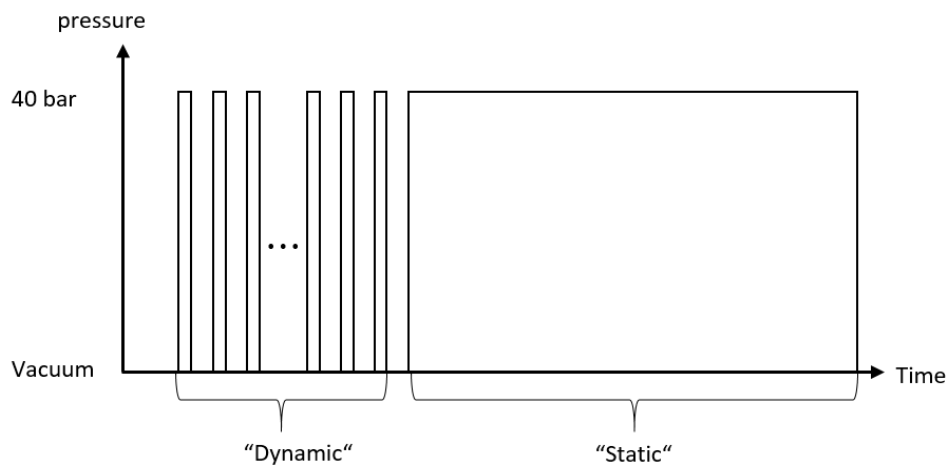

**Fig. 1.** Schematic diagram of the activation routine.

To investigate the effect of particle size on the activation process, the as-received TiFeMn sample was categorized, using manual sieves with a mesh size of 355 μm, 180 μm, 125 μm, and 90 μm (Fritsch GmbH), into four different particle size groups as presented in Table 1. For each measurement, the sample holder was evacuated for 2500 s at 70 °C and subsequently, the material was exposed to a hydrogen atmosphere of 40 bar at 70 °C. For comparison, the activation of the as-received TiFeMn (without sieving the particles) was also carried out under the same conditions.

**Table 1.** Sample designation of different particle size classes.

| Particle size | Sample designation |
| --- | --- |
| > 355 μm | PZ355- |
| 180-355 μm | PZ180-355 |
| 125-180 μm | PZ125-180 |
| 90-125 μm | PZ90-125 |
| < 90 μm | PZ-90 |
| As received (without sieving) | PZ-600 |

A Zeiss EVO MA10 (Carl Zeiss) Scanning Electron Microscope (SEM) was used for the morphological study on gold-sputtered samples (Secondary Electrons detector). Energy Dispersive X-ray Spectroscopy (EDX – Oxford Instruments) analysis data (elemental quantification and distribution maps) and SEM images collected with the Back-Scattered Electrons (BSE) detector (Carl Zeiss) were obtained on pristine samples (not sputtered). The measurements were performed at 20 kV with a working distance of 8.5 mm.

The morphology and the EDX-mapping of materials were characterized by a SEM (Cameca SX100 Zeiss GeminiSEM 500). To avoid moisture and/or oxygen contaminations during the sample preparation, a small amount of material was placed on a special sample holder inside a dedicated argon-filled glove box (< 1 ppm $O_2$ and $H_2O$). The sample holder was then evacuated before being transported to the SEM and was opened only after a high vacuum (50 Pa) had been reached inside the SEM chamber. The resolution of the images was set down to 1 μm. The elemental composition of the samples and their oxide layers was determined with a EDX detector operating in a FEI Nova 200 dual-beam SEM/FIB microscope. The sub-surface was revealed by cross-section milling with a focused (Ga+) ion beam (FIB) at an angle of 52° to the particle surface in the SEM. The EDX maps were done on the surface and exposed cross-sections.

The particle size distribution of the investigated specimens was determined by a Camsizer X2 instrument (Retsch Technology). The analysis generates a particle stream that is precisely characterized by an optical system; LED stroboscope light sources and two high-resolution digital cameras achieve a recording rate of over 300 images per second, which a powerful software evaluates in real-time.

The surface area was determined *via* the Accelerated Surface Area and Porosimetry System (ASAP, Micromeritics). To remove impurities from the sample before the measurement, the sample was heated between 80 °C to 120 °C under vacuum. Then the adsorption and desorption isotherms were measured with helium (gas quality 5.0) and a 9-point-BET analysis is performed to determine the surface area of the powder.

X-ray diffraction (XRD) analyses were carried out using a Bruker D8 Discover diffractometer (Bruker AXS GmbH) equipped with a Cu K$\alpha$ radiation ($\lambda$ = 1.54184 Å) X-ray source and a VÅNTEC-500 area detector; the images acquired by the 2D detector, each encompassing a range of more than 10°, were merged and integrated with DIFFRAC.EVA software. The diffraction patterns were acquired in nine steps in the $2\theta$ range from 10° to 110°, with an exposure time of 400 s per step and a step size of 10°. A small amount of powder was placed onto a sample holder and sealed with an airtight dome made of polymethylmethacrylate (PMMA). Rietveld refinement was carried out with the software Material Analysis Using Diffraction (MAUD) [44, 45]. The structure models of known phases were found in the International Crystal Structure Database (ICSD) via the software ICSD-Desktop.

X-ray Photoelectron Spectroscopy (XPS) technique was employed to examine the surface characteristics and chemistry of the as-received and the as-cycled TiFeMn materials. Therefore, the powder was pressed into pellets. This process was carried out by a hydraulic press, with a pressure of approx. 2000 bar. XPS measurements were performed using a KRATOS AXIS Ultra DLD (Kratos Analytical) equipped with a monochromatic Al K$\alpha$ anode working at 15 kV (225 W). For the survey spectra, pass energy of 160 eV was used, while for the region spectra the pass energy was 20 eV. The investigated area was a slit of 300 × 700 µm. Ar etching was performed with an etching rate of 7 nm/min related to $Ta_2O_5$ (acceleration voltage 3.8 kV with an extraction current of 160 µA) to clean the surface of the pellet sample. For all of the samples, no charge neutralization was necessary. The evaluation and validation of the data were carried out with the software CASA-XPS version 2.3.19. Calibration of the spectra was done by adjusting the C1s signal to 284.8 eV. For the region files, background subtraction (U 2 Tougaard) was performed before calculation.

The difference in thermal expansion behavior of the TiFe intermetallic and the oxide phase $Ti_4Fe_2O$ was properly assessed by first-principle calculations of their thermal properties. Vienna *ab initio* simulation package version 5.4.4 (VASP) [46, 47] was used to obtain the Hessian matrix employing density functional perturbation theory. The exchange-correlation effects were described by the revised version of the Perdew-Burke-Ernzerhof functional (RevPBE) [48] within the generalized gradient approximation. The calculations were carried out with a plane-wave energy cutoff of 480 eV and a k-point grid corresponding to a spacing of 0.2 Å$^{-1}$. An open-source package, Phonopy version 2.10.0 [49], was interfaced with VASP to calculate the thermal expansion under the quasi-harmonic approximation. The Helmholtz free energies of TiFe and the oxide at 12 different volumes (obtained by varying their equilibrium lattice parameters) were fitted to the Murnaghan equation of states (EOS) at temperatures from 0 to 800 K. Change in the volume dependence of the free energies with the temperature increase was evaluated to obtain the thermal expansion coefficients.

## 3. Results

### 3.1. Microstructure and morphology analysis

The graph in Fig. 2 shows the XRD patterns of the as-received and as-cycled TiFeMn materials. As shown in Fig. 2a, the as-received TiFeMn is composed of three different detectable phases: TiFe, $Ti_4Fe_2O$, and $Ti_{0.8}Fe_{0.2}$. According to the Rietveld refinement result (Fig. S1 in the Supplementary Information, SI), the relative amount of crystalline phases for TiFe, $Ti_4Fe_2O$, and $Ti_{0.8}Fe_{0.2}$ is 91.43 wt. %, 6.97 wt. % and 1.60 wt. %, respectively. Mn containing phase was not detected *via* XRD. The reason for this is that the Mn atoms substitute the Fe atoms in the TiFe solid solution. This substitution also leads to an increase in the cell size of TiFe [19]. After 15 absorption-desorption cycles, the XRD diffraction pattern of the material (Fig. 2b) presents the same phases found in the as-received material (TiFe and $Ti_{0.8}Fe_{0.2}$); moreover, the existence of $TiH_2$ ($\phi$) with small intensities was noticed, while $Ti_4Fe_2O$ cannot be detected in the cycled sample. The reason for this latter outcome is the possibility for $Ti_4Fe_2O$ (and mixed η-phase suboxides) to transform into a stable hydride, which release the hydrogen again only under high-temperature and low-pressure conditions [50]. Therefore, after cycling, the small amount of this ternary oxide drops below the detection threshold and it is likely accompanied by equally undetectable fractions of η-phases and their derived hydride, with wider peaks, depending on the hydrogenation level. According to Rietveld refinement (SI-Fig. S2), the phase composition of the TiFeMn material after cycling is 97.55 wt. % TiFe, 0.68 wt. % $Ti_{0.8}Fe_{0.2}$ and 1.77 wt. % $TiH_2$. The presence of $TiH_2$ may induce further cracks in the compound, facilitating the hydrogen sorption of the TiFeMn material upon cycling [41].

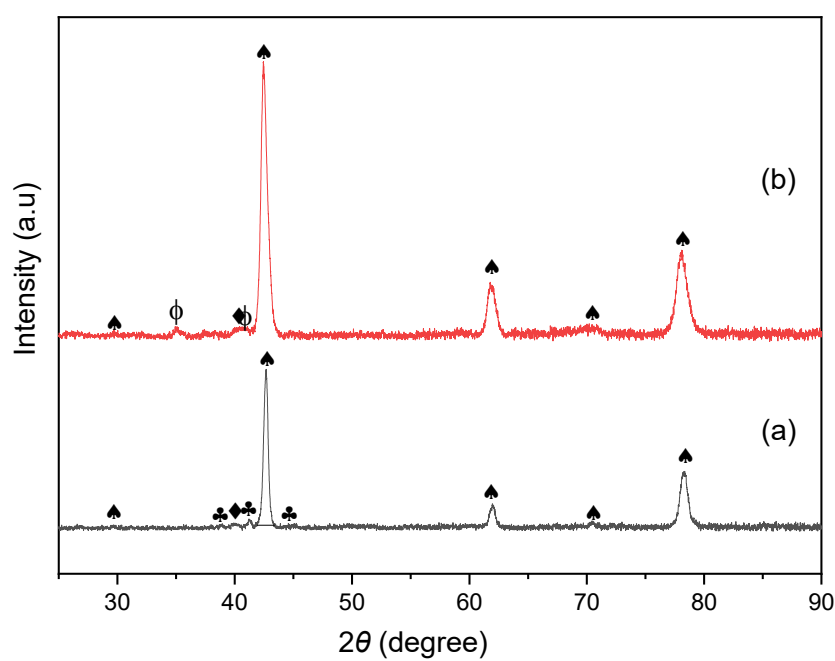

**Fig. 2.** XRD patterns of (a) as-received TiFeMn and (b) as-cycled TiFeMn.
♠ = TiFe, ♣ = $Ti_4Fe_2O$, ♦ = $Ti_{0.8}Fe_{0.2}$, ϕ = $TiH_2$

The micrograph in Fig. 3a shows a FIB image of a cross-section of a TiFeMn particle. The elemental mapping analysis of the region labeled "Cutting edge" and magnified in Fig. 3b, is shown in Fig. 3c-Fig. 3f. It can be seen in Fig. 3c that the oxygen content on the untreated surface (above the cutting edge) is higher (bright area) than that of the inner part, which is below the cutting edge (dark area). This is an indication of the presence of oxide phases on the surface of the sample. In contrast to the oxygen distribution, the other elements (Ti, Mn, and Fe) are homogeneously distributed within the material and show no difference between the surface and the inner area (above and below the cutting edge, respectively). According to the EDX quantitative analysis (SI-Fig. S3), the oxygen content is 7.71 at. %, while the atomic percentage of Ti, Fe, and Mn are 46.35 at. %, 42.05 at. % and 3.89 at. %, respectively.

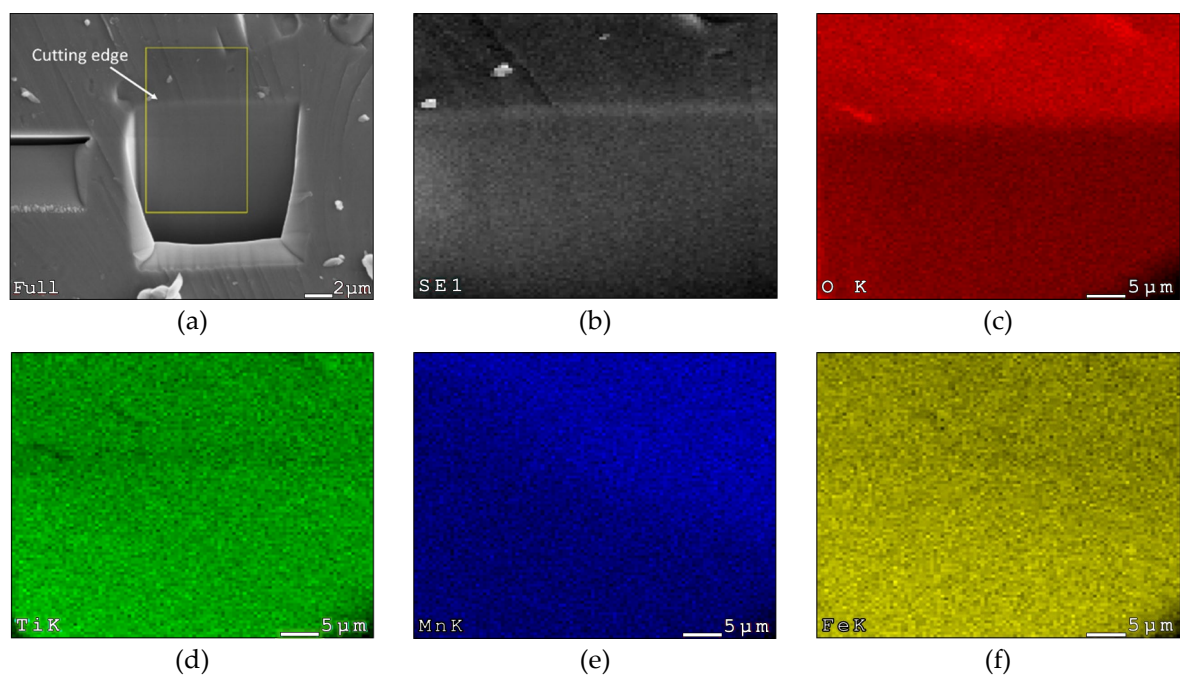

**Fig. 3.** SEM with FIB image and EDX mapping analysis of the as-received TiFeMn material: (a) FIB image; (b) SEM of the marked area from (a); (c)-(f) elemental mapping of oxygen, titanium, manganese, and iron, respectively.

The morphology of the TiFeMn sample at different stages (as-received, and as-cycled) was also characterized using SEM, as shown in Fig. 4. The particle size of the as-received TiFeMn material, displayed in Fig. 4a, varies in a wide range up to a few hundred micrometers and is significantly reduced after cycling. In fact, the particle size of the sample after cycling (Fig. 4b) is likely to be smaller than that of the as-received material. In conjunction with the formation of hydride in the frame of the activation process and the associated expansion of the material, the further fractures taking place in the bulk material are noticed for the cycled sample. The close-up view of cracks at a higher magnification level allows seeing the few thin cracks already existing in the as-received sample (Fig. 4c, Fig. 4e), while the cracks of the as-cycled sample (Fig. 4e, Fig. 4f) are bigger and appear with greater frequency.

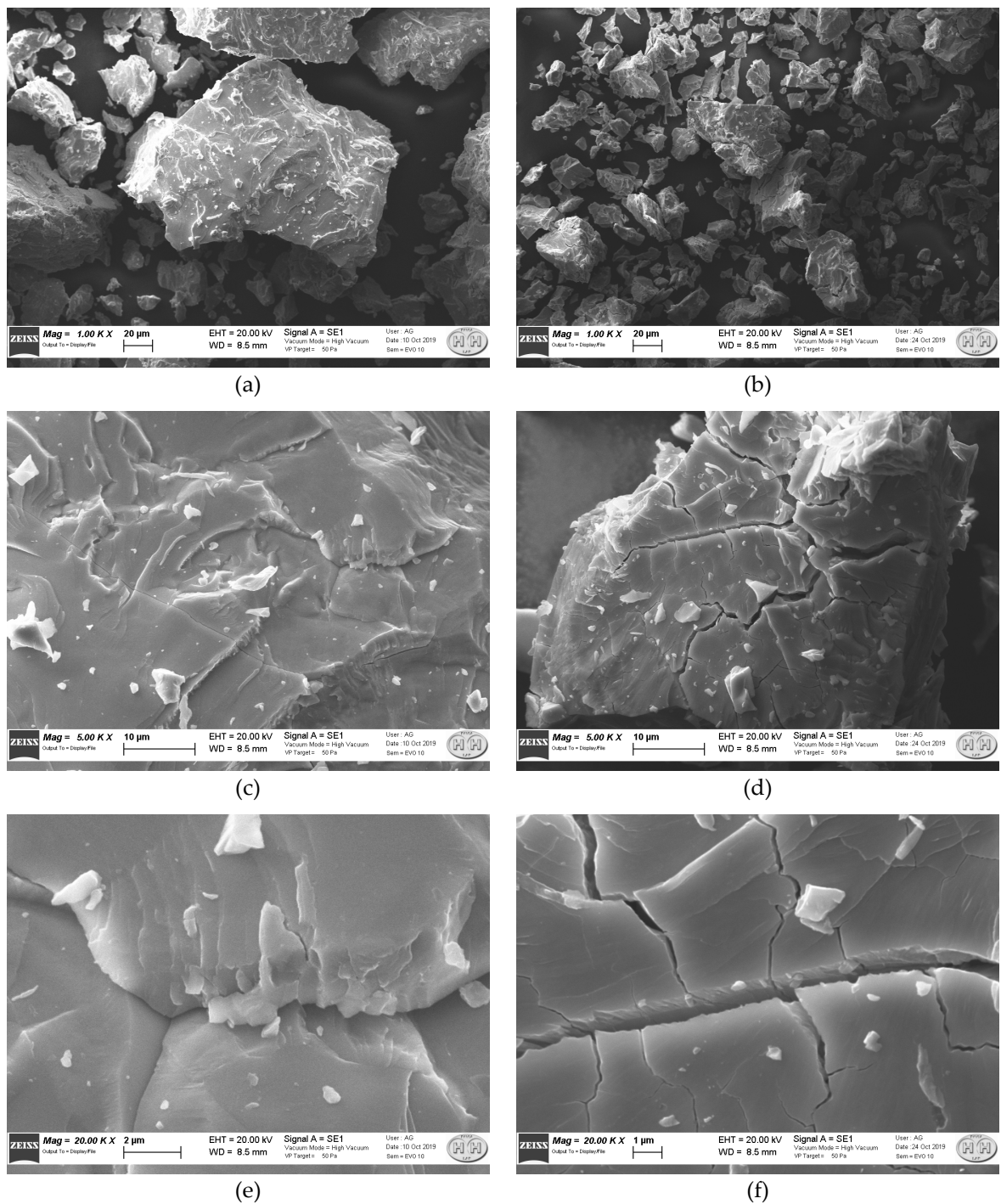

**Fig. 4.** SEM images at three different magnification levels acquired on the TiFeMn samples (a, c, e); and after 15 hydrogen absorption-desorption cycles (b, d, f).

The plots in Fig. 5 show the particle size distribution of the as-received and the activated TiFeMn powder. It is evident that the as-received powder has a wide range of particle sizes, up to 600 μm. Its size distribution (Fig. 5a) is homogeneous, with a large spreading range up to 441 μm, which represents almost 90 % of the particles. By activating and subsequent cycling 15 times, this value is reduced to 77 μm (Fig. 5b). These results are in good agreement with the reduction of particle size estimated by SEM analysis. As a consequence, the surface area of the as-cycled TiFeMn obtained by BET analysis is about 15.6 times higher than that of the as-received material, *i.e.* 0.5132 m$^2$/g versus 0.032 m$^2$/g.

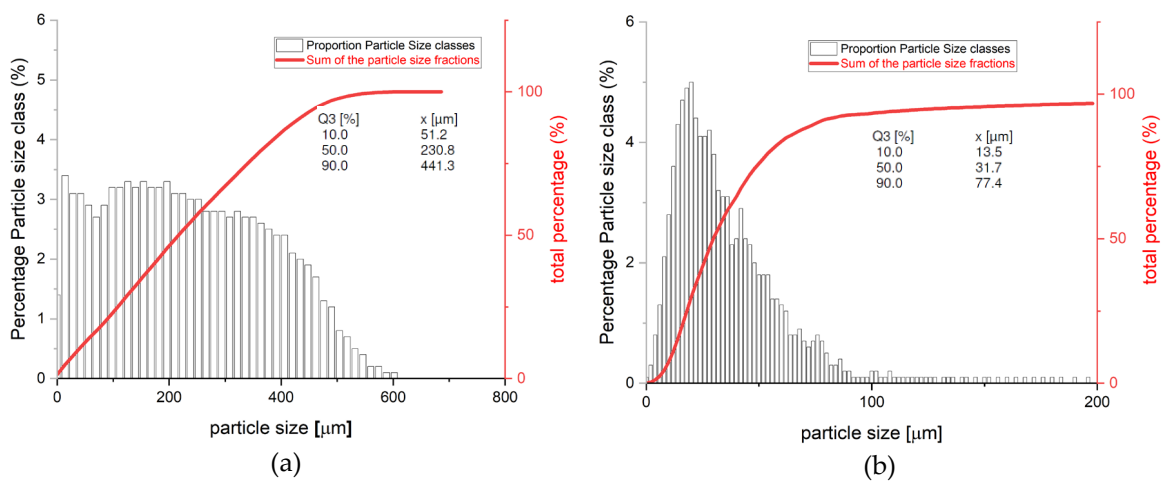

**Fig. 5.** Particle size distribution of the as-received and as-cycled TiFeMn material

In order to examine the surface of the TiFeMn material, the elemental composition on the near-surface of the pelletized sample was characterized by the XPS technique. Table 2 shows the XPS data of the TiFeMn pellets and their corresponding elemental compositions, before and after Ar etching. As shown in Table 2, the as-received TiFeMn contains 37.3 at. % O, suggesting the existence of high concentrations of oxides on the surface of the sample. The oxidation states of elements existing in this sample are shown in Table 2. In addition, contamination of impurities (*e.g.* Mg, C) can be found, which was probably caused by the production, storage, and handling processes. After 1800 s Ar etching, the concentration of Fe, Ti, and Mn rise, while the oxygen content is significantly reduced (Table 2). In fact, the oxygen concentration is reduced by a factor of 4 while the yield of Ti, Fe, and Mn increases by a factor of 5, 3, and 1.2, respectively. It is remarked that the rise of Mn content after Ar etching is very small in comparison to other elements (Ti, and Fe), implying that the oxide layer on the material surface may be dominated by Ti-, and Fe-containing species. The depth profile of the chemical composition over the Ar-ion bombardment (Fig. 6a) shows that the oxygen content gradually decreases by a factor of 4 after 180 s Ar etching, resulting in a penetration depth of approx. 21 nm with an Ar etching rate ~ 7 nm/min (related to $Ta_2O_5$). Further Ar etching does not further reduce the content of oxygen and magnesium and carbon contamination. Upon cycling, the binding energies and the oxidation states of the elements of the TiFeMn sample before and after Ar etching (SI-Fig. S4) are similar to those of the as-received material. That is to say, also in this case the atomic concentration of elements (Ti, Fe, and Mn) increases during the cleaning with Ar-ion bombarding, while the oxygen concentration is reduced by a factor of approx. 4 after 180 s Ar etching (Fig. 6b). However, the oxygen content of the cycled sample is higher than that of the as-received sample (41.8 at. % *versus* 37.3 at. %). It should be noted that the oxidation states for Ti could be changed by the bombardment with Ar-ions, nevertheless, it is evident that there is a high proportion of oxides on the surface of the TiFeMn material.

**Table 2.** Elemental compositions (at. %) of the as-received and the as-cycled TiFeMn pellets before and after Ar etching.

|  | As-received | | As-cycled | | |
| --- | --- | --- | --- | --- | --- |
| Elemental compositions | before Ar etching (at. %) | after Ar etching (at. %) | before Ar etching (at. %) | after Ar etching (at. %) | Oxidation states |
| O 1s | 37.3 | 9.8 | 41.8 | 11.9 | $O^{2-}$ |
| Ti 2p | 6.8 | 34.2 | 12.6 | 34.6 | $Ti^{4+}$ |
| Fe 2p | 15.7 | 48.4 | 20.9 | 46.8 | $Fe^{3+}$ |
| Mn 2p | 2.3 | 2.8 | 2.7 | 2.9 | $Mn^{2+}$ $Mn^{3+}$ |
| C 1s | 34.2 | 1.5 | 18.4 | 0.6 | - |
| Mg 2p | 3.8 | 3.2 | 3.6 | 3.2 | - |

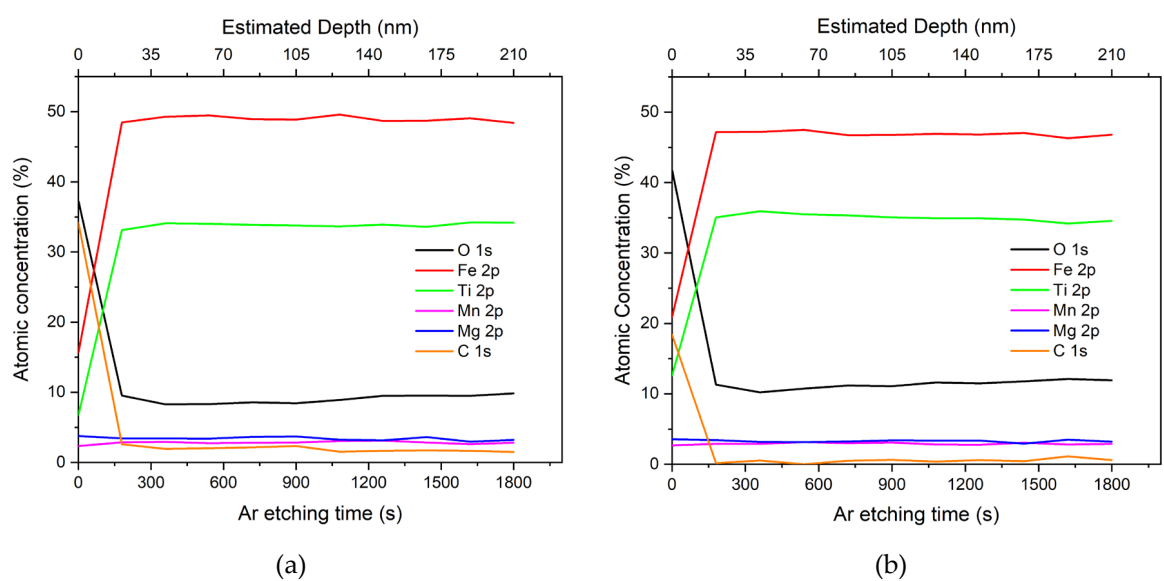

**Fig. 6.** Atomic concentration of the elements *versus* Ar etching time: (a) as-received and (b) as-cycled TiFeMn pellets.

### 3.2. Temperature dependence of the activation kinetics

To study the influence of temperature on the proposed activation process, 9 experiments were performed on the as-received TiFeMn sample at different "Dynamic" and "Static" temperatures, using the Sievert's apparatus. The first hydrogen loading profiles are displayed in Fig. 7 together with the corresponding incubation times of the materials at different temperatures. The incubation time is defined from the beginning of the "Static" period until the visible absorption of material observed in the first absorption curve (Fig. 7a). In Fig. 7a, it can be observed that at 50 °C "Dynamic" and 50 °C "Static" (denoted as 50 °C/50 °C sample), the sample shows a static incubation time of about 277 h, before the hydrogen absorption is initiated. Increasing the temperature of the "Static" phase to 70 °C (denoted as 50 °C/70 °C sample) and even further to 90 °C (denoted as 50 °C/90 °C sample), the incubation time of the material is significantly reduced by a factor of 2.8 and 15.8, respectively. It is also noticed that the activation is possible at all the temperatures investigated (from 50 °C to 90 °C) and that an increase of temperature in the "Static" phase from 50 °C to 90 °C considerably reduces the total activation time of the material, while the amount of hydrogen absorbed by the material decreases accordingly, from 1.78 wt. % $H_2$ to 1.45 wt. % $H_2$, as observed in Fig. 7b. The more pronounced capacity reduction detected in all the three routines ending with a 90 °C "Static" phase is explained by the reduced extent and the possible slope of the plateau at that temperature [19]; at 90 °C, the hydride forming at 40 bar H₂ is in lower amount, but this does not infer permanent loss of capacity.

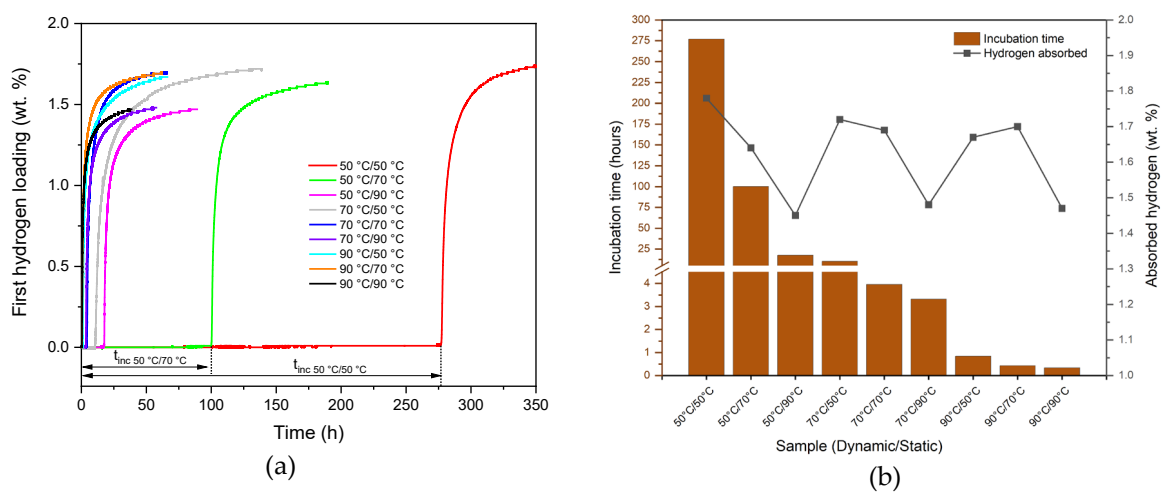

**Fig. 7.** (a) First hydrogen loading profiles of the as-received TiFeMn material and exemplary the incubation times $t_{inc\ 50\ °C/50\ °C}$ and $t_{inc\ 50\ °C/70\ °C}$ for the tests 50 °C/50 °C and 50 °C/70 °C and (b) corresponding incubation time of material and quantity of absorbed hydrogen for all samples, measured at different "Dynamic" and "Static" temperatures under 40 bar of hydrogen.

Similarly, increasing the temperature of the "Dynamic" phase from 50 °C (50 °C/50 °C sample) to 70 °C (70 °C/50 °C sample) and further to 90 °C (90 °C/50 °C sample), while keeping the same temperature of the "Static" phase at 50 °C, the incubation times of the material becomes shorter. These intervals for the 70 °C/50 °C and the 90 °C/50 °C sample fall to 10.42 h and 0.84 h, which are steadily reduced by a factor of 26.6 and 330, respectively, compared to the 50 °C/50 °C sample. Besides, the absorbed hydrogen quantity slightly declines from 1.78 wt. % for the 50 °C/50 °C sample to 1.72 wt. % for the 70 °C/50 °C sample and to 1.67 wt. % for the 90 °C/50 °C sample (Fig. 7b). Comparing the temperature changes in both the "Dynamic" and "Static" periods versus the incubation time, it is remarked that an increase in the "Dynamic" temperature had a greater influence on the incubation times of the material than that of the "Static" temperature, *i.e.* the incubation times of the 70 °C/50 °C and the 50 °C/70 °C sample are reduced to 10.42 h and 100.1 h, compared to the 50 °C/50 °C sample (277 h). At the same "Dynamic" temperature of 90 °C with different "Static" temperatures of 50 °C (90 °C/50 °C sample), 70 °C (90 °C/70 °C sample), and 90 °C (90 °C/90 °C sample), the material is quickly activated, in less than 1 h for all the procedures considered (Fig. 7b). Particularly, the activation process of the 90 °C/70 °C and 90 °C/90 °C samples require only 25 and 20 min, respectively, suggesting that nearly no incubation time might be required beyond 100 °C. The results displayed above indicate that the 70 °C/70 °C procedure allows reaching a compromise in terms of performance for high hydrogen storage capacity, fast activation kinetics, technical effort, and the required energy for heating the material. For this reason, the 70 °C/70 °C procedure was used for further investigations.

To identify the rate-limiting step, which controls the overall reaction rate of the first hydrogen absorption process, the gas-solid models [51] were applied (SI-Table S1). The fitting procedure can be found in the literature [52,53] and has been successfully applied to metal-hydride systems. The hydrogen absorption data obtained after the material started the reaction following the 70 °C/70 °C procedure, as shown among other curves in SI-Fig. S5, were used to make a linear fit of the different model equations, by using the least-squares method. The fitting results of these solid-gas models are shown in Fig. 8; it can be seen that the D3 – three-dimensional diffusion model presents a good fitting agreement, with the highest correlation coefficient $R^2$ of 0.99395. This model is a support for further hypotheses, when discussing this metal hydrogen system, but does not consider all the factors involved.

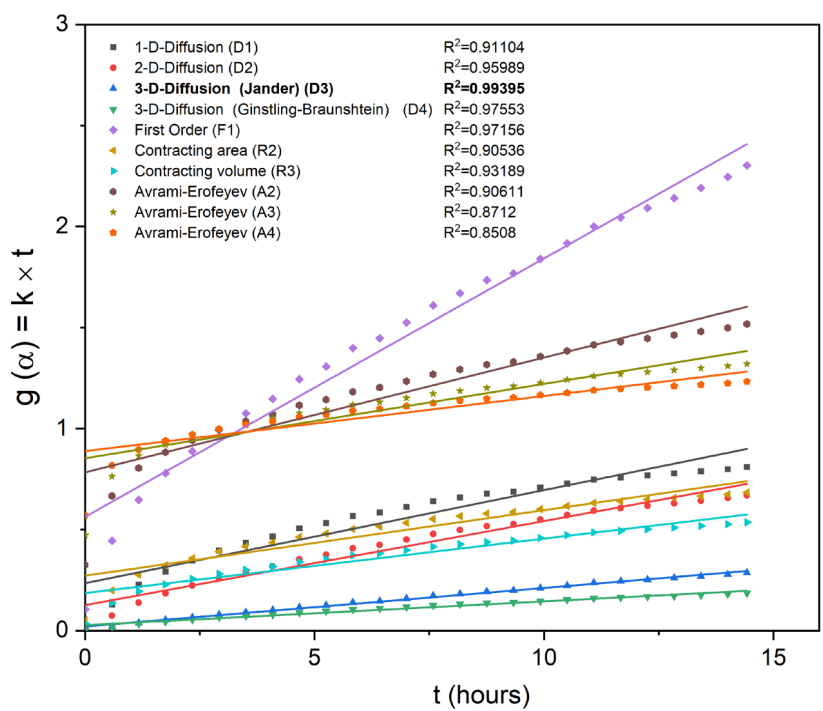

**Fig. 8.** Plots of integral forms of the solid-state models versus time for the as-received TiFeMn sample at the 1st hydrogenation at 70 °C, after the incubation time at 70 °C and under 40 bar of $H_2$. Linear fittings and the relative correlations coefficients are also reported.

To study the influence of temperature on the size of the metal lattice, the thermal expansion of TiFe and the oxide $Ti_4Fe_2O$, which are most abundant in the alloy in percentage terms, was calculated. Therefore, Fig. 9 shows the volume change and the thermal expansion as a function of temperature.

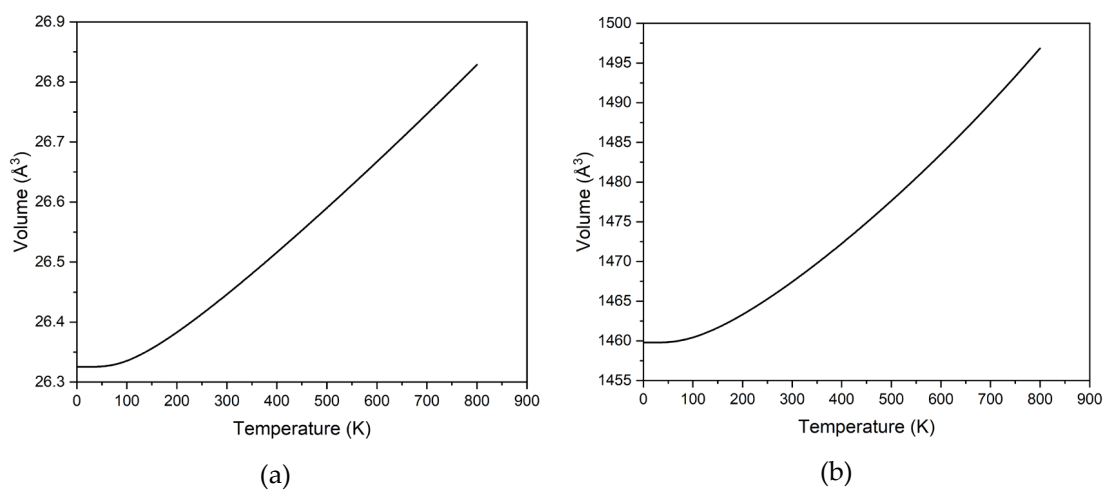

(a)     (b)

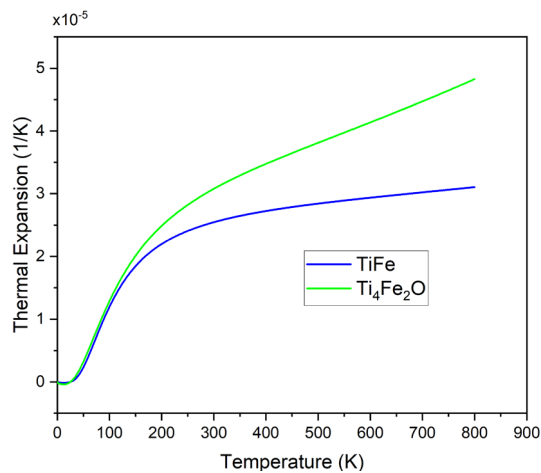

(c)

**Fig. 9.** Volume expansion of the lattice of (a) TiFe and (b) Ti$_4$Fe$_2$O and the thermal expansion of TiFe and Ti$_4$Fe$_2$O (c) as functions of temperature.

In Fig. 9, the thermal expansion of TiFe and Ti$_4$Fe$_2$O phases from 0 to 800 K was calculated. It can be seen that the volumetric thermal expansion coefficient and the cell volume of both TiFe and Ti$_4$Fe$_2$O increase with the increase of temperature. Due to the temperature increase from 50 °C to 90 °C, the cell volume of TiFe rises from 26.4619 Å$^3$ at 50 °C to 26.4898 Å$^3$ corresponding to an increase of 0.1054 %. The calculated values are close to the experimentally determined volume of 26.2510 Å$^3$ for TiFe [15]. The volumetric thermal expansion coefficient of TiFe increases from 2.5957·10$^{-5}$ K$^{-1}$ at 50 °C to 2.6683·10$^{-5}$ K$^{-1}$ at 90 °C. In comparison, the cell volume of Ti$_4$Fe$_2$O increases from 1468.4926 Å$^3$ to 1470.4095 Å$^3$, which represents an increase of 0.1305 %. The thermal expansion coefficients of Ti$_4$Fe$_2$O increase from 3.1796·10$^{-5}$ K$^{-1}$ at 50 °C to 3.3400·10$^{-5}$ K$^{-1}$ at 90 °C.

### 3.3. Particle size dependence of the incubation time

It is well known that the activation is also strongly affected by the particle size of the material; in the scientific literature, smaller particles were often found to have faster kinetics during the initial hydrogen absorption compared to larger particles, as evidenced by similar findings [54]. In the present work, the effect of particle size on the activation of TiFeMn material was also investigated. Fig. 10a shows the first hydrogenation of TiFeMn samples with different particle sizes. Apparently, the sample with the largest particles (PZ355-) has the lowest incubation time and shows the fastest hydrogenation kinetics, for example, it starts to absorb hydrogen after a 0.72 h incubation and finally reaches a total capacity of 1.79 wt. % H$_2$ after 24 h loading. For comparison, the PZ180-355 sample starts to load hydrogen after 2.33 h of incubation, but it has a higher hydrogen capacity of 1.88 wt. % after 24 h loading. The highest capacity amongst the investigated samples (about 1.9 wt. %) was obtained for the PZ180-355, after 29.77 h loading. The TiFeMn particles in the range of 90-125 µm (PZ90-125) presented the most static incubation time of 100 h. In addition, its hydrogen absorption kinetics is slower than the one of the other particle size classes and can store 1.61 wt. % H$_2$ after 24 h loading. From the aforementioned results, it can be summarized that the activation of the material is faster with the increase of its particle size and the incubation time required, for the investigated materials, follows a regular trend: 355 µm > 180-355 µm > 125-180 µm > 90-125 µm (Fig. 10b). It is also noticeable that the incubation time for the as-received TiFeMn sample containing unsorted particles up to 600 μm is even shorter than that of the samples with particles lower than 180 μm. The sample with a particle size lower than 90 μm could not be activated under these conditions with the given time of 200 hours.

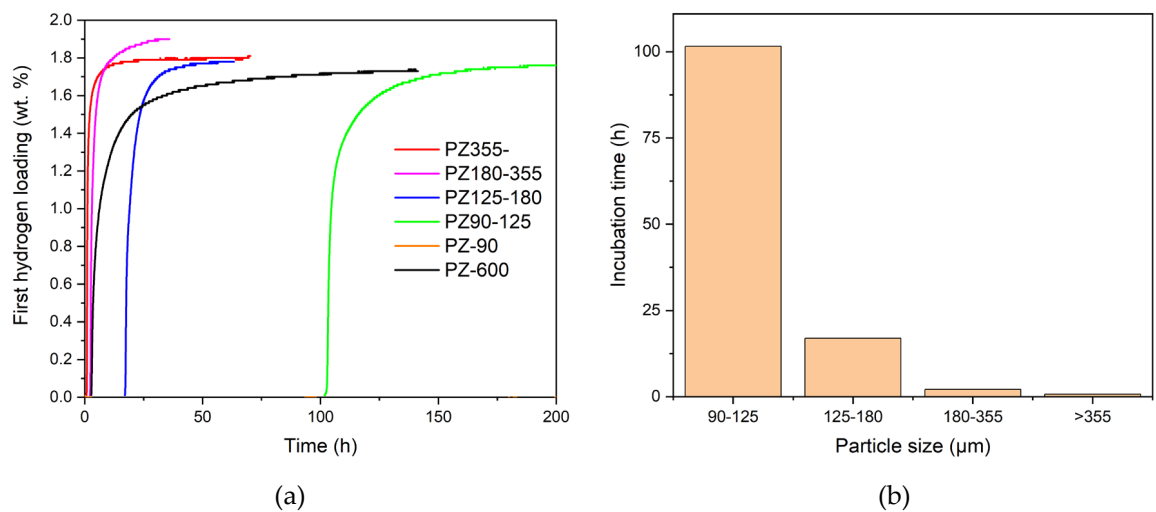

**Fig. 10.** (a) First hydrogen loading of TiFeMn with different particle sizes and (b) corresponding incubation times measured at 40 bar and 70 °C.

*3.4. Particle size dependence of the reaction kinetics and the cyclability*

Since the activation method proposed in this study takes place at relatively low temperatures, *i.e.* below 100 °C, it is assumed that only surface adsorbed oxygen and impurities are removed during the dynamic phase. The actual oxide layer is neither removed nor significantly affected, as this occurs only at higher temperatures [16] and the material still contains oxides after being activated. Moreover, as reported above, the particle size influences the initial activation of the TiFeMn material. Hence, at this point, it was important to check if the particle size of the material influences the reaction kinetics and its reversibility.

The cycling measurements were carried out under isothermal conditions (70 °C at 40 bar of hydrogen pressure). Fig. 11 shows the cycling profiles of the PZ90-125 (Fig. 11a), the PZ125-180 (Fig. 11b), and the PZ-600 (Fig. 11c) after the activation procedure and subsequent desorption/re-absorption cycles. By comparing Fig. 11a with Fig. 11b, it can be seen that the hydrogen absorption properties of both samples PZ90-125 and PZ125-180 are alike. During the first 600 s, all the absorption kinetic curves have nearly the same rate and reach a hydrogen storage capacity of about 1.12 wt. %. The achievable capacity for the second absorption of the material is 1.32 wt. % $H_2$, and this value decreases over cycling, *i.e.* from 1.32 wt. % to 1.19 wt. % after 15 cycles. For comparison, the cycling tests of the sample PZ-600 were also analyzed. As seen in Fig. 11c, there are also no significant differences in the following absorption cycles of sample PZ-600, except that in the second absorption curve, the maximum hydrogen storage capacity is about 1.2 wt. %, a value lower than those of the PZ90-125 and the PZ125-180 samples. The total hydrogen storage capacity reaches roughly 1.15 wt. %, corresponding to a loss of 0.05 wt. % $H_2$ over 15 cycles for the PZ-600 sample, which is lower than the others. The results obtained from this investigation indicate that the difference in particle size of the material does not affect significantly the kinetic properties of the absorption-desorption cycles performed after the activation.

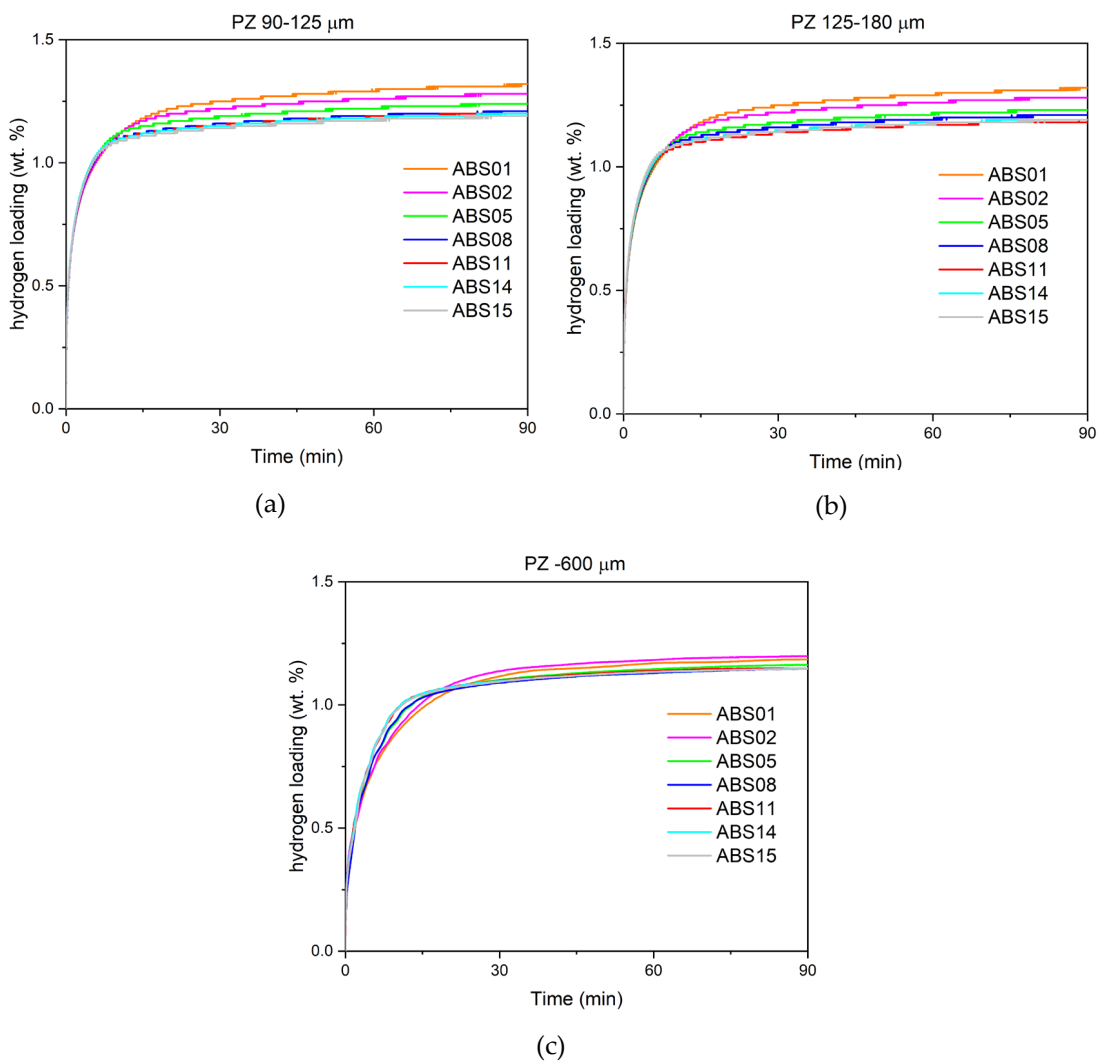

**Fig. 11.** Cycling profiles of the samples: (a) PZ90-125, (b) PZ125-180, and (c) PZ-600, after the initial hydrogen absorption-desorption cycle, measured under the isothermal condition at 70 °C and 40 bar hydrogen.

## 4. Discussion

The results reported in this study show a possibility to activate the TiFeMn material at temperatures below 90 °C using the proposed activation method. Following the tendency of the low-temperature experiments, it is believed that the activation should be possible even at temperatures below 50 °C, although the incubation period might take longer. The repetition of vacuum and hydrogen refilling during the "Dynamic" period should primarily have a cleaning effect to remove adsorbed oxygen and other impurities, but also weaken the expected oxide layer on the particle surface especially at higher temperatures. According to the volumetric measurements (Fig. 7), the activation is more effective at higher temperatures than at lower ones, for example, the incubation time of the 90 °C/90 °C sample is less than 1 hour, while 3.96 hours and 277 hours are necessary for the 70 °C/70 °C and the 50 °C/50 °C samples, respectively. This indicates that the incubation period could be completely avoided with the further increase of the temperature. In particular, the increase in temperature of the "Dynamic" phase is observed to be more beneficial than that of the "Static" phase, providing a reason to explain the much faster activation of the 90 °C/50 °C sample compared to the 50 °C/90 °C one.

The surface investigation, performed *via* XPS, shows that a layer of oxides has formed on the near-surface of the TiFeMn particles. The thickness of this layer can be estimated at 21 nm (Fig. 6), which is in good agreement with the literature, where the oxide layers in TiFe were found to have a thickness between 20 and 30 nm [16,55]. This layer is composed of different Ti and Fe oxides (Table 2) and might serve as a diffusion barrier, thus inhibiting the penetration of hydrogen into the bulk. This assumption is supported by the kinetic modeling (Fig. 8), in which the rate-limiting step of the first hydrogen loading kinetic curve is three-dimensional diffusion-controlled growth. The content of Ti, Fe, Mn, and O reported in Table 2 remained constant until the end of the XPS measurement. Therefore, the continuous presence of oxygen could be primarily ascribed to the fact that the powder used for the XPS examination was pressed into pellets and every single particle has an oxide layer over the whole surface. The increase of oxygen content over cycling, noticeable in the XPS analysis of the cycled sample, could be explained by the availability of newly created surfaces, which sensibly react with the low amount of unavoidable residual oxygen during the handling. The results deriving from this surface investigation suggest that the activation at temperatures lower than 90 °C does not fully affect and/or reduce the oxide layers. In fact, the removal of the oxide layer can be achieved only at high temperatures, as reported in the literature [42, 56].

Moreover, the cell expansion due to the substitution of Mn in the TiFe structure could make the activation easier, besides decreasing the equilibrium pressure of the first plateau [15, 19]. The enhanced activation of Mn-substituted TiFe alloy is, indeed, related to these changes in the microstructure and to the precipitation of small clusters of atoms and/or segregation of metal atoms at the grain boundaries that markedly increase the boundaries reactivity and could locally influence the composition at the surface [7, 16]. In addition, fractures on the surface of the as-received material (Fig. 4c) and the inhomogeneous distribution of the alloy constituents should help the hydrogen atoms being able to enter the bulk at low temperatures. As the temperature increases, the $Ti_4Fe_2O$ oxide and the TiFe bulk material underneath might expand with different strains, which can cause stresses in the material. The different expansions of the lattice volume of TiFe and $Ti_4Fe_2O$, shown in Fig. 9 with an increase of 0.1054 % and 0.1305 %, support this assumption, but with these small different percentages this effect should not be the determining factor for the accelerated activation.

For this reason, the effect of the temperature on the diffusion of hydrogen is calculated. Since the oxide layer consists of, among others, Fe and Ti oxides (see Table 2), the diffusion coefficient of $TiO_2$-oxide is considered. By increasing the temperature from 50 °C to 90 °C the diffusion coefficient increases from $2.0391 \cdot 10^{-21}$ m$^2$/s to $6.1203 \cdot 10^{-20}$ m$^2$/s which corresponds to an increasing factor of about 30. These coefficients were calculated by using the 0-dimensional temperature-dependent diffusion model from [57]. Compared with experiments 50 °C/50 °C and 50 °C/90 °C (see Fig. 7), for example, an acceleration of the incubation time by a factor of about 16 can be seen, due to the increase in temperature from 50 °C to 90 °C in the "Static"-phase, which is quite close to the previous one. The difference between the two factors could be due to the presence of other oxides and impurities in addition to the $TiO_2$-oxide. In fact, suboxides deriving from $Ti_4Fe_2O$ and with an analogous structure (η-phases), are also known to combine with hydrogen, therefore increasing their cell volume [7, 50, 58]. This phenomenon can be listed among the reasons for the presence of fractures, as well, exposing surface in the particles (Fig. 12); an experimental confirmation may be found in the absence of marked

Ti$_4$Fe$_2$O peaks in the XRD pattern of the cycled sample (Fig. 2b), suggesting a conversion to other compounds.

These diffusion coefficients of TiO$_2$ are smaller by a factor of about 10$^{-5}$ compared to the value of TiFe with the range of 1.1467·10$^{-15}$ m$^2$/s at 50 °C to 8.2862·10$^{-15}$ m$^2$/s at 90 °C, which is calculated also by the diffusion equation from [59]. This comparison is also a confirmation that the oxide layer existing on the particle surface constitutes a diffusion barrier. In addition, these values are also smaller than the diffusion coefficients of β-Ti with the order of 10$^{-14}$ m$^2$/s and other metallic hydrides in order of 10$^{-12}$ m$^2$/s up to 10$^{-10}$ m$^2$/s [7]. These results makes it clear that the improved diffusion, caused by the temperature increase in particular, is mainly responsible for the acceleration of the activation. Assuming that the oxide layer in the fractures, which are already present in the as-received material, is thinner or more permeable (see Fig. 12), these features should additionally accelerate the activation.

Apart from the temperature, the particle size also influences the activation kinetics of the material. In fact, the incubation period of the sample with larger particles is shorter than in the case of smaller particles and the reaction kinetics of the first hydrogen loading decreases with decreasing particle size (Fig. 10a). Interestingly, the initial hydrogen loading curve of the as-received sample (PZ-600) seems to be a weighted average of the other curves for different particle sizes. The incubation time of the PZ-600 sample is very close to that of the middle-sized particles (PZ180-355) and both their reactions proceed at the same rate, at the beginning of the first hydrogenation process. Afterward, the slope of the curve reduces considerably and the curve characteristics are getting closer to those of the PZ90-125 sample, which are characterized by slower kinetics and a lower total capacity. This is probably due to the fact that every as-received sample contains particles with a size range up to 600 μm, thus, in the same sample, the larger particles would be activated earlier than the smaller particles (see Fig. 10a). This can also be explained by the fact that the exothermic reaction increases the temperature in the sample, to such an extent that the particles in direct contact with the active ones are also directly activated, like in a chain reaction. This assumption is confirmed by the results of the first hydrogen loading experiments, which have shown that an increase in temperature of the "Dynamic" period from 50 °C to 90 °C reduces the incubation time considerably. The particles with a size lower than 90 μm seem to be not activated at all and, as a consequence, they inhibit the activation of the bigger particles. This could be the reason why the PZ-600 sample has a lower final capacity than the other particle sizes. The reason for the easier activation of the larger particles may also be due to the improved diffusion caused by the temperature increase. This causes the hydrogen concentration in the lattice to increase more rapidly until the first hydride formation occurs. The resulting pressure forces on the layer from the inside then lead to the breakup of the oxide-layer, allowing a strongly accelerated penetration of hydrogen. Due to the higher volume to surface ratio of the large particles, these inner forces are stronger than those of the small particles, leading to the formation of more pronounced cracks and new reaction surfaces and subsequently promoting the faster reaction kinetics. Furthermore, this larger ratio of volume, with respect to surface area, could lead to the large particles having a higher number and more pronounced cracks and fractures already before the activation process, which additionally accelerate the activation of the larger particles (see Fig. 12).

In the framework of the kinetics experiments from Fig. 11, it can be determined that the particle size after activation has no significant influence on the reaction kinetics. The reduction in the final capacity in the course of the cycling can be explained by the fact that the powder is crushed by expansion and shrinkage of the metal lattice, decreasing the ratio of volume to surface area of the particles and, therefore, slightly lowering the number of available sites for hydrogen. In addition, some hydrogen atoms could be bound in the form of a stable binary hydride ($TiH_2$), which could be formed during hydrogenation.

In addition, the inhomogeneous distribution of constituent elements of the TiFeMn and contaminations, like oxides, can also influence the activation process. One possibility is that, as previously reported [60], an increase of the Ti content in the oxide layer significantly accelerates the activation of the material, even at lower temperatures. Furthermore, oxides like $Ti_4Fe_2O$ could probably act as an activation promotor and easily form hydrides [42], making the activation under 100 °C possible, or possibly act also as a hydrogen transfer catalyst [39,61]. Such inhomogeneity can easily occur in industrial production on larger scales. The suggested mechanism for the formation of the cracks and defects can be seen in Fig. 12.

During the activation, the formation of metal hydride and the associated lattice expansion cause the particles to decrepitate. After multiple hydrogen absorption-desorption cycles, the TiFeMn material undergoes the reduction of particle size, leading to a higher reaction surface (the specific surface area of the material, calculated with the BET method, was increased by a factor of 15.6 after cycling) and larger cracks/fractures are formed with greater incidence (Fig. 4d, Fig. 4f). These fractures and defects allow the hydrogen to penetrate the metal lattice easily and quickly from the fresh TiFeMn surfaces and therefore to avoid the incubation period, resulting in faster sorption kinetics in the following hydrogen absorption-desorption cycles (Fig. 11).

However, the increase of cycles presents the mentioned disadvantage of a slightly reduced capacity, especially for the smaller particles, as noticeable in Fig. 11a and b. Besides the aforesaid particle decrepitation, the sequential expansion and shrinkage of the cells in the crystal lattice, respectively during hydrogenation and dehydrogenation, generates internal stresses and defects, which are proven to affect the characteristics of the pressure-composition isotherms. Typically, their hysteresis gap is reduced, but the pressure for the second plateau is gradually increased [6, 7]. This implies that, at constant measuring pressure, the driving force to form the γ-phase is lowered by cycles, until the concentration of lattice strain is stable. Then, the final hydrogen capacity is expected to become constant, as suggested in Fig. 11c for a broader distribution of particles, enclosing lower strained lattices.

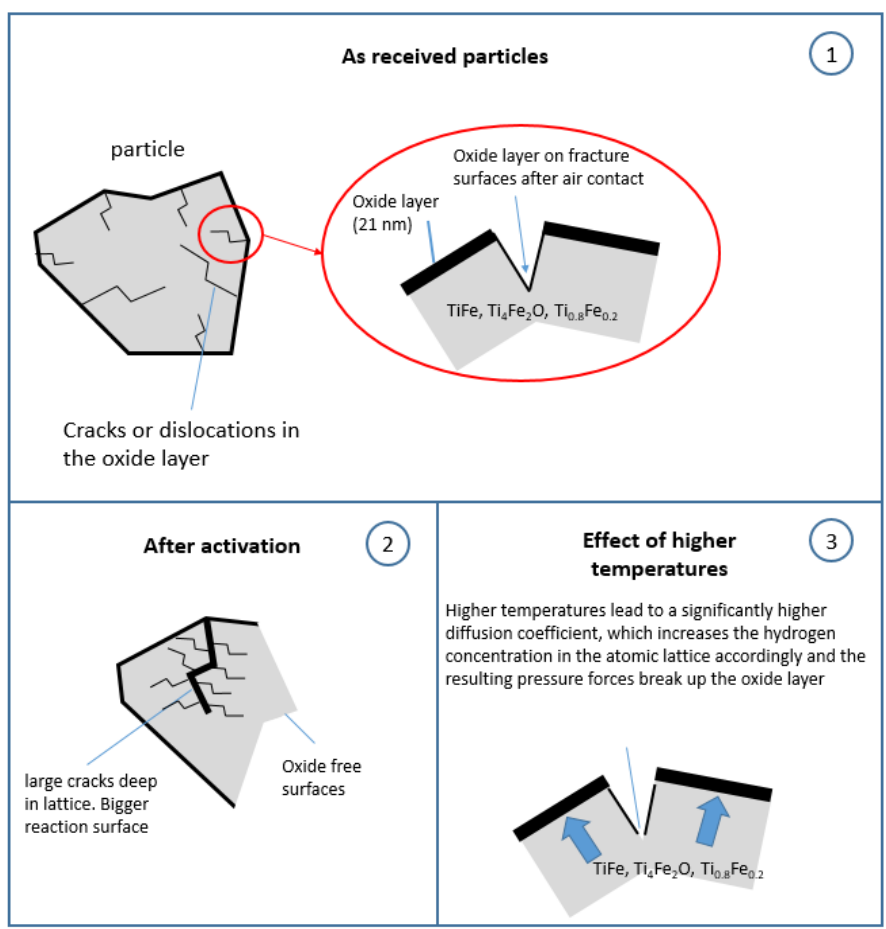

**Fig. 12.** Schematic diagram of cracks and defects in the as-received particle surface, the effect of the activation on the particles and the Effect of higher temperatures on the activation process.

## 5.   Conclusions

An effective routine for the activation of industrially produced TiFeMn powders, aiming at low costs and low technical efforts, was proposed. Within the scope of this study, the following results were obtained:

- Several experiments reveal the possibility of activating the samples in all the investigated temperature range (between 50 °C and 90 °C).
- An approximately 21 nm thick oxide layer consisting of different Fe and Ti oxides was observed on the particles of the as-received material; this layer might act as a diffusion barrier to prevent the penetration of hydrogen into the lattice.
- An increase in temperature of the activation routine strongly promotes diffusion of the hydrogen through the oxide-layer and thus accelerates the activation process. This effect could be also improved by cracks and defects already existing on the particle surface.
- The better diffusion of hydrogen leads to a faster increase in the hydrogen concentration in the lattice and thus to an earlier hydride-formation. The resulting inner forces against the oxide-layer break this layer and improve the penetration of further hydrogen in the bulk material.
- Due to the lower surface-to-volume ratio, the larger particles have a higher number of fractures on the particle surface and build up internal stresses, caused by the hydride formation, that are greater than in the smaller particles and that break the oxide layer more effectively and earlier in the process.
- The activation itself promotes further cracks and defects in the material (especially in the near-surface layers), leading to the formation of new clean surfaces, which facilitate the hydrogen diffusion into the lattice and finally improve the kinetics of the following sorption cycles.
- Other possible factors, which influence the activation, depend on the inhomogeneity of the alloying elements.

All these considerations are indicative of the statistical characteristics of the activation, which depends mainly on the size and number of cracks and defects in the particles. By observing that the temperature-dependent diffusion of hydrogen and the cracks or defects are mainly responsible for hydrogen to overcome the oxide barrier, it can be concluded that the mechanical effect in the activation method would be more likely to activate the TiFeMn alloy at low temperatures. However, considering the use of large batches of industrially produced powders, the proposed method is a promising alternative, as it avoids handling and storing the powders outside the intended vessel and it could be performed directly in the storage system itself, owing to the mild conditions.

**Acknowledgment**

This work was partially funded from the Fuel Cells and Hydrogen 2 Joint Undertaking (now Clean Hydrogen Partnership) under grant agreement No. 826352, HyCARE project. The Joint Undertaking receives support from the European Union's Horizon 2020 research and innovation programme, Hydrogen Europe, Hydrogen Europe Research and Italy, France, Germany, Norway, which are all thankfully acknowledged.
This research paper is partially funded by dtec.bw – Digitalization and Technology Research Center of the Bundeswehr.
The work was also supported by a doctoral scholarship of the Hochschule Bonn-Rhein-Sieg, University of Applied Sciences.
Further thanks go to Maximilian Passing and Thomas Breckwoldt, for their support with experiments.

**Declaration of Competing Interest**

The authors declare that they have no known competing financial interests or personal relationships that could have appeared to influence the work reported in this paper.

**CRediT authorship contribution statement**

D.M. Dreistadt: Methodology, Validation, Formal analysis, Investigation, Visualization, Data Curation, Writing - Original Draft, Writing - Review & Editing. T.-T. Le: Methodology, Formal analysis, Visualization, Writing - Original Draft, Writing - Review & Editing. G. Capurso: Conceptualization, Validation, Formal analysis, Visualization, Data Curation, Writing - Original Draft, Writing - Review & Editing, Supervision, Project administration. J.M. Bellosta von Colbe: Writing - Review & Editing, Supervision. A. Santhosh: Methodology, Software, Formal analysis, Visualization, Writing - Original Draft. C. Pistidda: Conceptualization, Resources, Writing - Review & Editing, Supervision. N. Scharnagl: Investigation, Resources. H. Ovri: Investigation, Resources. C. Milanese: Investigation, Resources, Writing - Review & Editing. P. Jerabek: Conceptualization, Software, Formal analysis, Supervision, Writing - Review & Editing. T. Klassen: Writing - Review & Editing, Project administration, Funding acquisition. J. Jepsen: Conceptualization, Resources, Writing - Review & Editing, Supervision, Project administration, Funding acquisition.

**Appendix A. Supporting information**

Supplementary data associated with this work can be found in the attached Supplementary information file (and in the published version at https://doi.org/10.1016/j.jallcom.2022.165847). Dataset related to this article can be found at https://doi.org/10.5281/zenodo.6623361.

# An Effective Activation Method for Industrially Produced TiFeMn Powder for Hydrogen Storage

David Michael Dreistadt [a], Thi-Thu Le [a], Giovanni Capurso [a,1]*, José M. Bellosta von Colbe [a], Archa Santhosh [a], Claudio Pistidda [a], Nico Scharnagl [b], Henry Ovri [c], Chiara Milanese [d], Paul Jerabek [a], Thomas Klassen [a,e], Julian Jepsen [a,e]

[a] Institute of Hydrogen Technology, Helmholtz-Zentrum hereon GmbH, Max-Planck-Straße 1, 21502 Geesthacht, Germany
[b] Institute of Surface Science, Helmholtz-Zentrum hereon GmbH, Max-Planck-Straße 1, 21502 Geesthacht, Germany
[c] Institute of Materials Mechanics, Helmholtz-Zentrum hereon GmbH, Max-Planck-Straße 1, 21502 Geesthacht, Germany
[d] Pavia Hydrogen Lab, C.S.G.I. & Chemistry Department, University of Pavia, viale Taramelli 16, 27100 Pavia, Italy
[e] Helmut Schmidt University, Holstenhofweg 85, 22043 Hamburg, Germany

* Corresponding author: giovanni.capurso@hereon.de

# Supplementary Information

---

[1] Present address: Polytechnic Department of Engineering and Architecture, University of Udine, via del Cotonificio 108, 33100 Udine, Italy

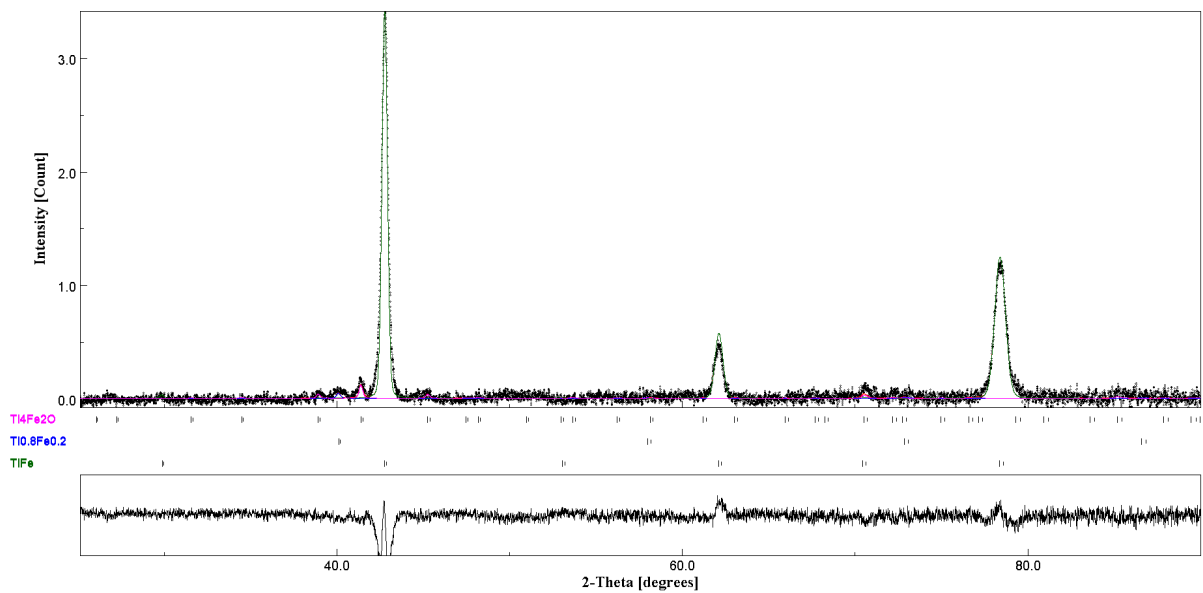

**Fig. S1.** XRD patterns of the as-received TiFeMn sample (black line) and Rietveld refinement ($Ti_4Fe_2O$ – purple line, $Ti_{0.8}Fe_{0.2}$ – blue line, and TiFe – green line); the weighted profile R-factor ($R_{wp}$) of the refinement is 62 %.

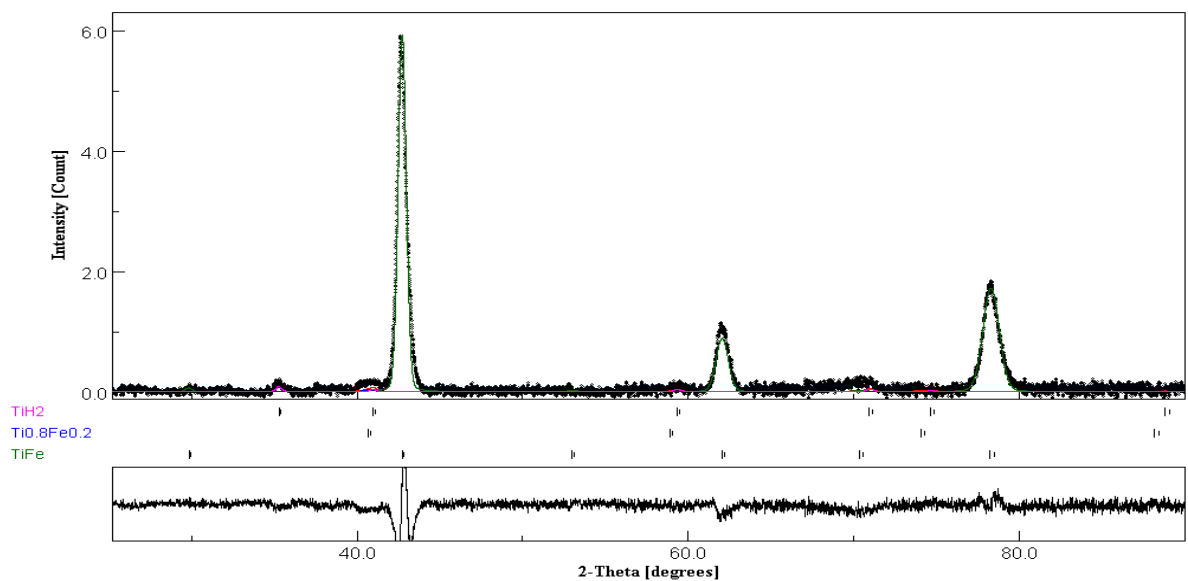

**Fig. S2.** XRD patterns of the as-cycled TiFeMn sample (back line) and Rietveld refinement ($TiH_2$ – purple line, $Ti_{0.8}Fe_{0.2}$ – blue line, and TiFe – green line); the $R_{wp}$ index of this refinement is 56 %.

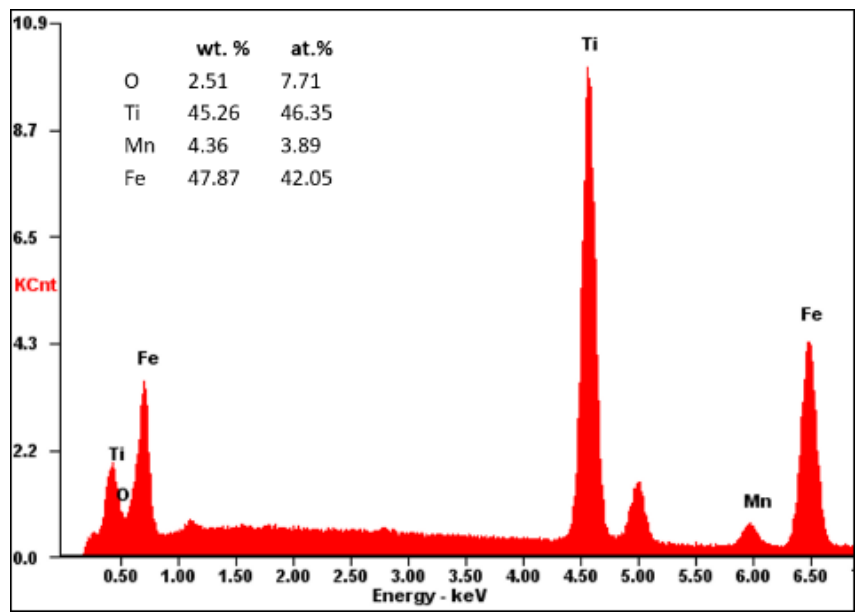

**Fig. S3.** EDX quantitative composition analysis of the as-received TiFeMn material.

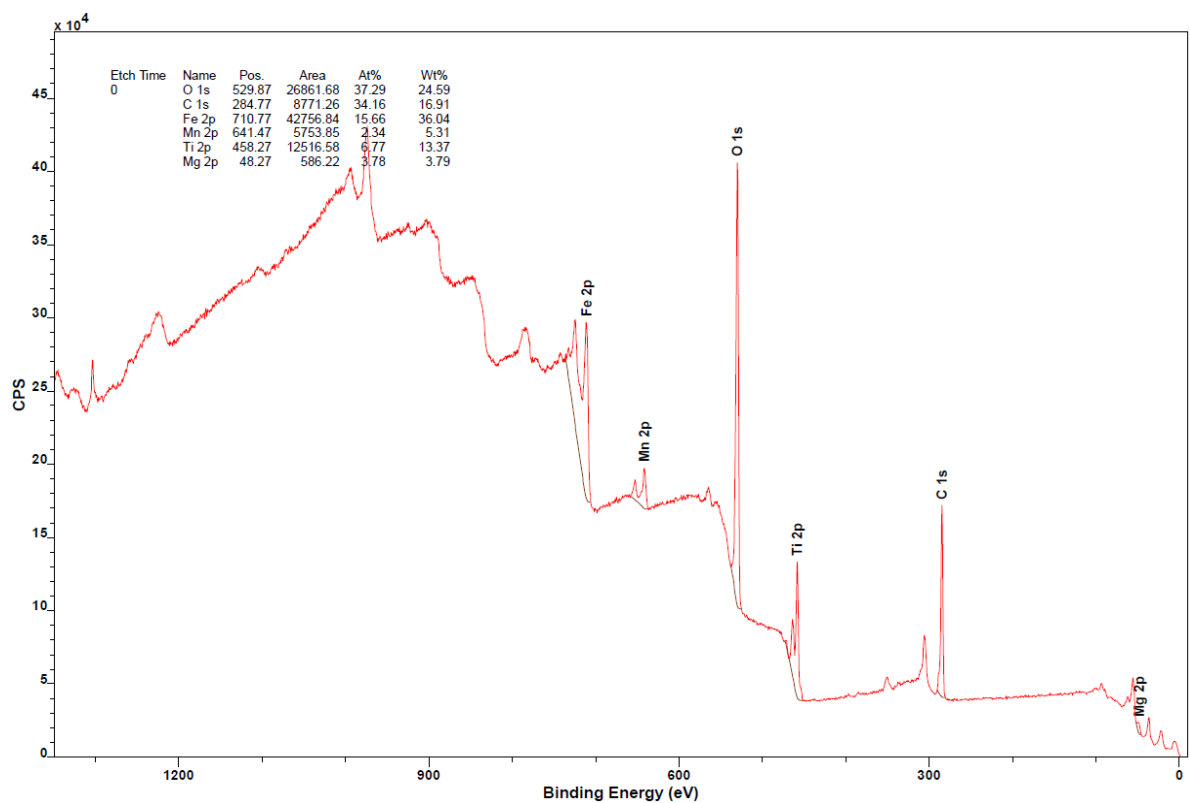

(a)

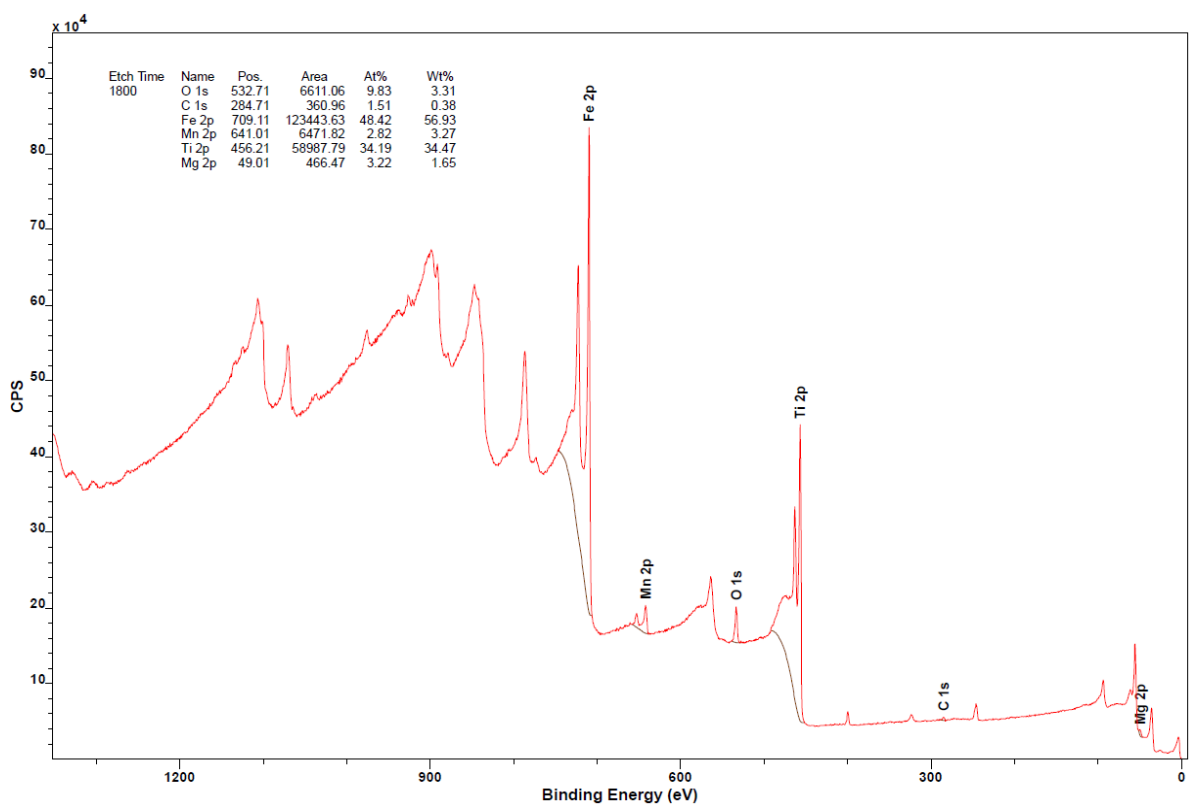

(b)

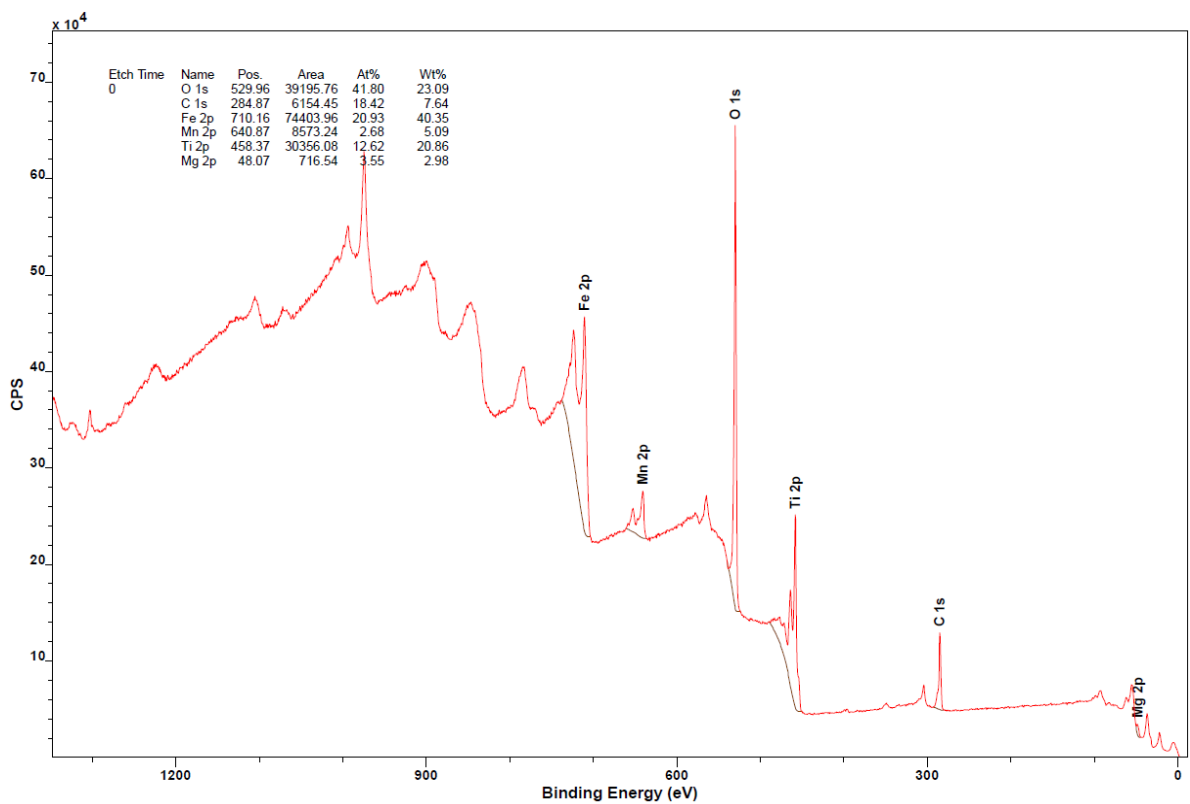

(c)

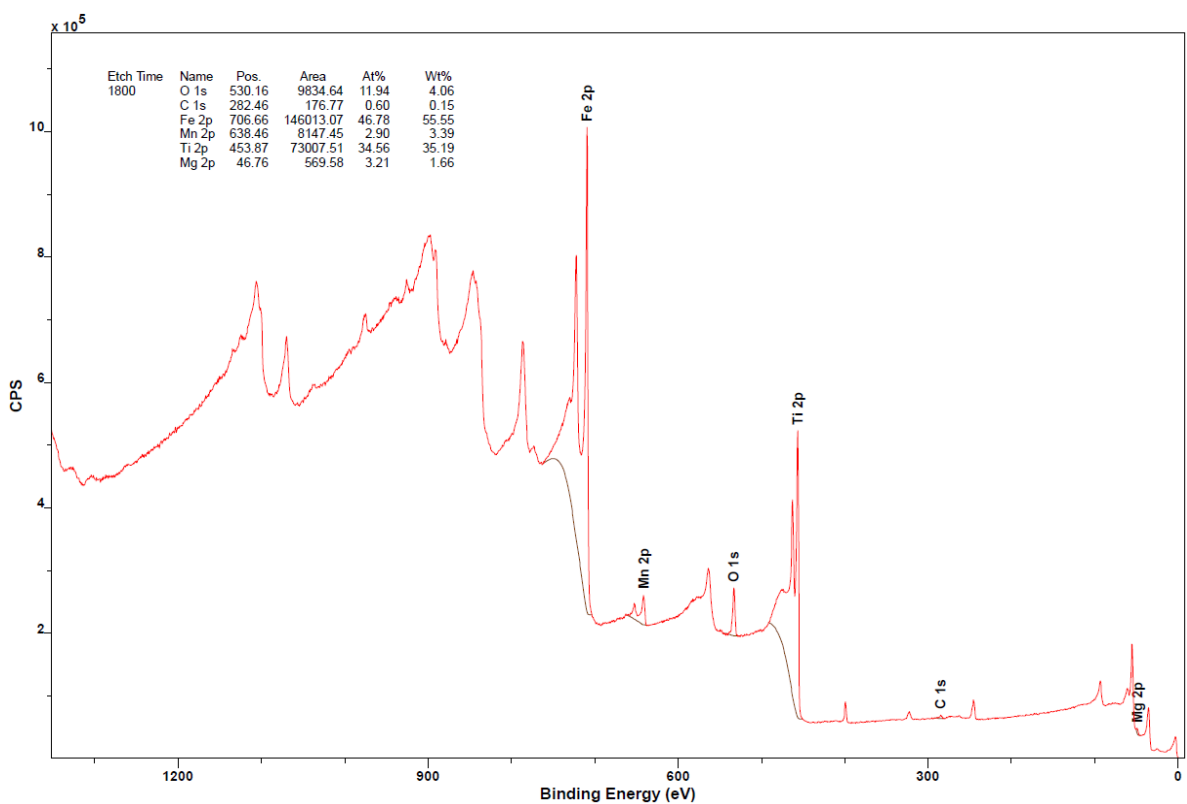

(d)

**Fig. S4.** XPS spectra of TiFeMn pellets: as-received TiFeMn before (a) and after Ar etching (b); as-cycled TiFeMn before (c) and after Ar etching (d).

**Table S1.** Summary of integral expressions for different reaction models.

| Kinetic model | | Differential form $f(\alpha) = 1/k \, d\alpha/dt$ | Integral form $g(\alpha) = kt$ |
|---|---|---|---|
| Nucleation-growth models | Power law, P2 | $2\alpha^{1/2}$ | $\alpha^{1/2}$ |
| | Power law, P3 | $3\alpha^{1/2}$ | $\alpha^{1/3}$ |
| | Power law, P4 | $4\alpha^{1/2}$ | $\alpha^{1/4}$ |
| | Avrami-Erofeyve, A2 | $2(1-\alpha)[-\ln(1-\alpha)]^{1/2}$ | $[-\ln(1-\alpha)]^{1/2}$ |
| | Avrami-Erofeyve, A3 | $3(1-\alpha)[-\ln(1-\alpha)]^{2/3}$ | $[-\ln(1-\alpha)]^{2/3}$ |
| | Avrami-Erofeyve, A4 | $4(1-\alpha)[-\ln(1-\alpha)]^{3/4}$ | $[-\ln(1-\alpha)]^{3/4}$ |
| | Prout-Tompkins, B1 | $\alpha(1-\alpha)$ | $\ln[\alpha/(1-\alpha)]+c*$ |
| Geometrical contraction models | Contracting area, R2 | $2(1-\alpha)^{1/2}$ | $1-(1-\alpha)^{1/2}$ |
| | Contracting area, R3 | $3(1-\alpha)^{2/3}$ | $1-(1-\alpha)^{1/3}$ |
| Diffusion models | 1-Dimensional Diffusion, D1 | $1/(2\alpha)$ | $\alpha^2$ |
| | 2-Dimensional Diffusion, D2 | $1-[1/\ln(1-\alpha)]$ | $((1-\alpha)\ln(1-\alpha))+\alpha$ |
| | 3-Dimensional Diffusion, D3 | $[3(1-\alpha)^{2/3}/[2(1-(1-\alpha)^{1/3})]$ | $(1-(1-\alpha)^{1/3})^2$ |
| | Ginstling-Brounshtein, D4 | $3/[2((1-\alpha)^{-1/3}-1)]$ | $1-(2/3)\alpha-(1-\alpha)^{2/3}$ |
| Reaction-order models | Zero-order (F0/R1) | $1$ | $\alpha$ |
| | First-order, F1 | $(1-\alpha)$ | $-\ln(1-\alpha)$ |
| | Second-order, F2 | $(1-\alpha)^2$ | $[1/(1-\alpha)]-1$ |
| | Third-order, F3 | $(1-\alpha)^3$ | $(1/2)[(1-\alpha)^{-2}-1]$ |

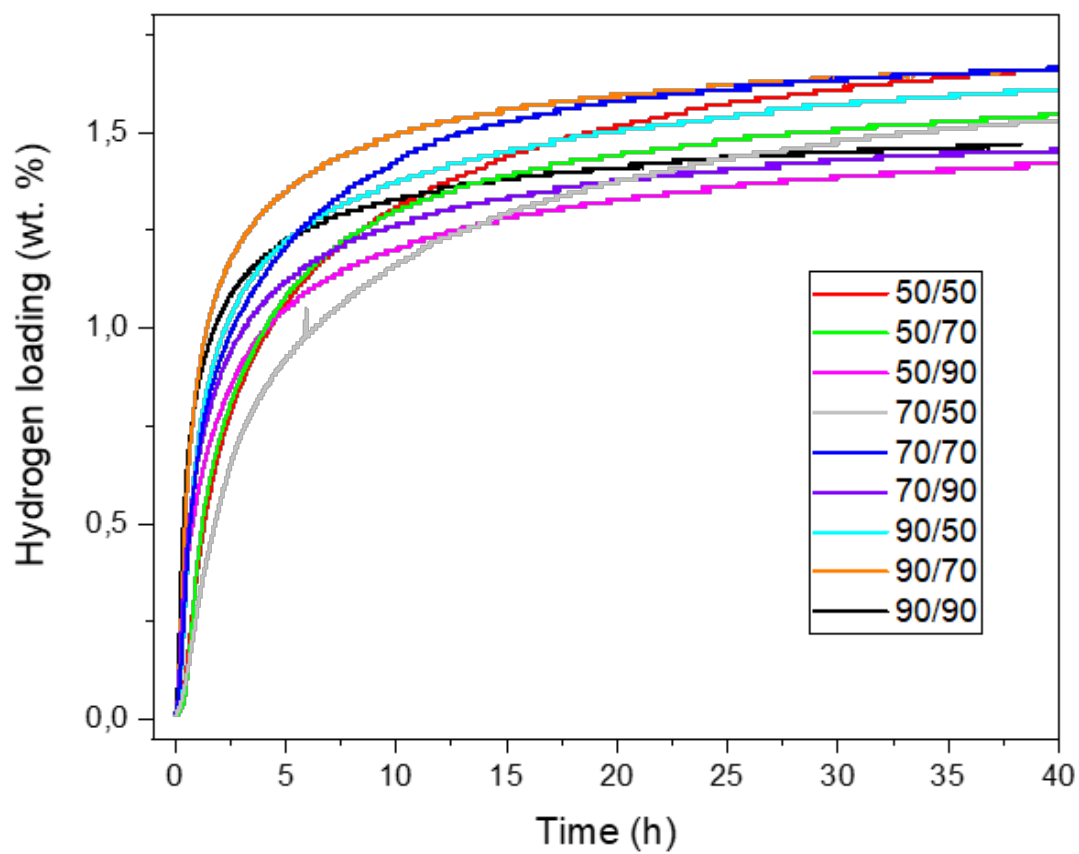

**Fig. S5**. Hydrogen loading profiles of the as-received TiFeMn material, measured at different dynamic temperatures and different static temperatures under 40 bar of hydrogen.